\documentclass[showpacs,prb,twocolumn]{revtex4}

\usepackage{amsfonts}
\usepackage{graphicx}
\usepackage{amsmath}
\usepackage{amssymb}
\usepackage{latexsym}
\usepackage{psfrag}

\usepackage[bookmarks, colorlinks=true, plainpages = false,citecolor 
= blue, urlcolor = black, filecolor = blue]{hyperref}

\begin{document}
\title{Nonlinear conductance quantization in graphene ribbons}
\author{S. Ihnatsenka}
\affiliation{Department of Physics, Simon Fraser University, Burnaby, British Columbia, Canada V5A 1S6}
\author{G. Kirczenow}
\affiliation{Department of Physics, Simon Fraser University, Burnaby, British Columbia, Canada V5A 1S6}

\begin{abstract}
We present numerical studies of non-linear conduction in graphene nanoribbons when a bias potential is applied between the source and drain electrodes. We find that the conductance quantization plateaus show asymmetry between the electron and hole branches if the potential in the ribbon equals the source or drain electrode potential and strong electron (hole) scattering occurs. The scattering may be at the ends of a uniform ballistic ribbon connecting wider regions of graphene or may be due to defects in the ribbon. We argue that, in ribbons with strong defect scattering, the ribbon potential is pinned to that of the drain (source) for electron (hole) transport.  In this case symmetry between electron and hole transport is restored and our calculations explain the upward shift of the conductance plateaus with increasing bias that was observed experimentally by Lin {\em et al}. [Phys. Rev. B \textbf{78}, 161409(R) (2008)]. 
\end{abstract}

\maketitle

\section{Introduction}
   
Conductance quantization in graphene nanoribbons was recently observed experimentally by Lin {\em et al.}\cite{Lin08} Unlike in the ballistic quantum wires in which conductance quantization had been observed previously,\cite{Kirczenow} the quantized conductance steps observed by Lin {\em et al.}\cite{Lin08} were two orders of magnitude lower than the conductance quantum $2e^2/h$. The observation\cite{Lin08} of such low quantized conductance values was surprising. However, it was subsequently shown theoretically\cite{disorder, adsorbate} that it can arise from enhanced electron backscattering near subband edges of disordered graphene ribbons with carbon atom vacancies\cite{disorder} or chemisorbed atoms or molecules.\cite{adsorbate} More recently, conductance quantization has also been reported in low conductance ribbons that were fabricated in a different way,\cite{Lian2010} and ballistic conductance quantization in graphene nanoconstrictions has also been observed.\cite{Tombros2011} With increasing bias voltage applied between source and drain electrodes the quantized conductance plateaus were observed by Lin {\em et al.}\cite{Lin08} to shift upward for both electron and hole conduction. It was suggested\cite{Lin08} that this effect was due to the increase with increasing bias of the number of subbands that are present in the energy window between the source and drain electrochemical potentials and therefore contribute to transport. However, whether this mechanism can {\em by itself} explain the experimentally observed non-linear transport effect\cite{Lin08} has not been investigated theoretically and is still an open question. In this paper we investigate non-linear conduction in graphene ribbons with the help quantum transport simulations. We find that strong electron scattering by defects and a specific type of potential profile in the ribbon under bias are both necessary for the the nonlinear transport phenomenon observed by Lin {\em et al.}\cite{Lin08} to occur.

We describe the models of the graphene ribbons and potential profiles that we study in Section \ref{Model}. We present our results for ballistic ribbons having uniform widths and having wide-narrow-wide geometries in Section \ref{Ballistic ribbons}. Our results for ribbons with defects are presented in Section \ref{Results} and the physics of the potential profiles in such ribbons is discussed in Section \ref{Discussion}. Our conclusions are summarized in Section \ref{Conclusions}. 

\section{Model}
\label{Model}

We describe graphene ribbons by the standard tight-binding Hamiltonian on a honeycomb lattice,
\begin{equation}
  H = \sum_i\epsilon_ia_i^{\dag }a_i - \sum_{\left\langle i,j\right\rangle} t_{ij}\left( a_i^{\dag}a_j + h.c. \right),
  \label{eq:hamiltonian}
\end{equation}
where $\epsilon_{i} = V_{bi}$ is the on-site energy that accounts for bias potential $V_b$ at site $i$; $t_{ij}=t=2.7$ eV is the matrix element between nearest-neighbor atoms. This Hamiltonian is known to describe the $\pi$ band dispersion of graphene well at low energies.\cite{Reich02} Spin and electron interaction effects are disregarded in present study. The effect of disorder is represented by bulk vacancies that are introduced by randomly removing carbon atoms and setting appropriate hopping elements $t_{ij}$ to zero. It is quantified by the probability of the carbon atoms being removed $p$, which is normalized relative to the whole sample. We refer the reader to our previous papers, Ref. \onlinecite{disorder,adsorbate}, and other studies\cite{Evaldsson08, Mucciolo09, Areshkin07} for the effects disorder of other types on conduction in graphene ribbons. 

\begin{figure}[t]
\includegraphics[keepaspectratio,width=\columnwidth]{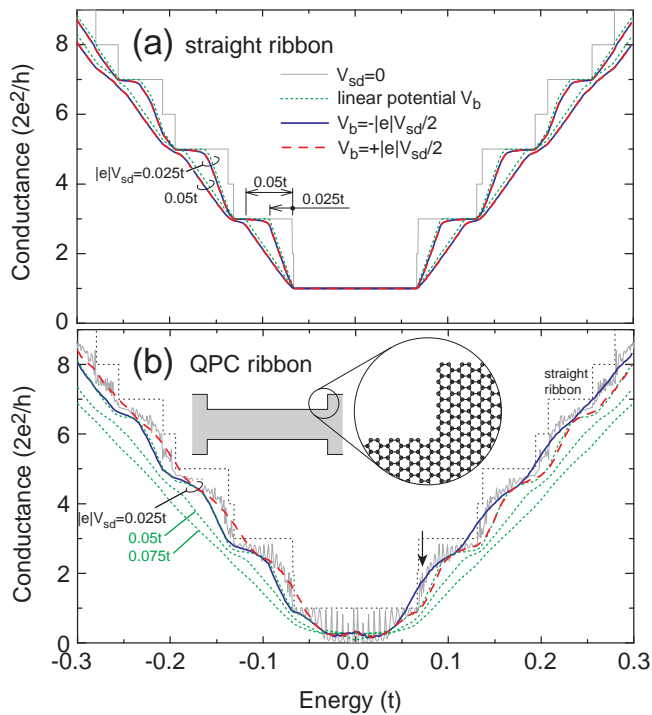}
\caption{(color online) Conductance as a function of the Fermi energy for ideal uniform (a) and QPC (b) graphene ribbons. The gray solid lines show zero bias conductance $V_{sd}=0$; the green dotted lines correspond to linear potential ramp along the structure; the solid blue and dashed red lines correspond to the entire potential drop $V_{sd}$ occurring at the source or drain end, respectively, of the ribbon in (a) and constriction in (b). The gray dotted line in (b) shows the conductance of the uniform ribbon for comparison; it is the same as the gray solid line in (a). The width and length are 10 nm and 100 nm for the ribbon in (a) and constriction in (b); the leads in (b) are twice as wide. The edges are armchair except for the wide-narrow boundaries in (b) that are shown in the inset. The arrow in (b) marks the Fermi energy for the DOS plots in Fig. \ref{fig:2}. $t=2.7$ eV.}
\label{fig:1}
\end{figure}

Electron transport out-of-equilibrium is modeled by application of a source-drain voltage $V_{sd}$ that biases the  electrochemical potentials of the source and drain leads as $\mu^s=+\frac{1}{2}\left| e \right| V_{sd} + E_\mathrm{F}$ and $\mu^d=-\frac{1}{2}\left| e\right| V_{sd}+ E_\mathrm{F}$, respectively, where $E_\mathrm{F}$ is the Fermi energy. Note that throughout this paper our definitions of the source and drain are such that {\em electrons} flow from the source to the drain. The potential profiles within experimentally realized graphene ribbons are not well understood since in addition to the source-drain and gate voltages they depend on the presence of charged adsorbates on the graphene, charged defects in the substrate and other uncharacterized materials properties. We will therefore consider three basic model potential profiles that might be realized in experimental devices and compare the results of our quantum transport calculations for these model profiles with the experimental results of Lin {\em et al.}\cite{Lin08} The profiles that we will consider are: a linear ramp of $V_b$, and constant $V_b$ along the channel with a potential drop either at the source or drain electrode. We will show here that only potential profiles of the latter type are consistent with the nonlinear transport data of Lin {\em et al.}\cite{Lin08} and will discuss the physical origin of such potential profiles in Section \ref{Discussion}.  For a non-zero $V_{sd}$ the electric current is calculated as\cite{Davies_book} $I=\frac{2e}{h}\int_{\mu_d}^{\mu_s} dE\;T(E, V_{sd})$ with $T(E, V_{sd})$ being the transmission coefficient. Here and below we assume zero temperature. In order to calculate $T(E, V_{sd})$ we employ the non-equilibrium Green's function method,\cite{Datta_book, Igor08} which is also used for calculation of the density of states (DOS) in the ribbon. The conductance of the ribbon is given by the Landauer formula\cite{Datta_book}
\begin{equation}
  G = \frac{I}{V_{sd}} = \frac{2e}{h}\frac{1}{V_{sd}} \int_{\mu^d}^{\mu^s}dE \; T(E, V_{sd}).
  \label{eq:conductance}
\end{equation}
For graphene ribbons with vacancies we also present the averaged conductance $\left\langle G\right\rangle$ that was obtained by averaging over 10 samples with different realizations of disorder.

\section{Ballistic ribbons}
\label{Ballistic ribbons}

Let us consider first the non-linear conductances of ideal graphene ribbons of finite length $L$ with the source-drain potential applied between its two ends. The latter are attached to semi-infinite leads that serve as electron reservoirs and have constant electrochemical potentials. The bias potential is allowed to drop inside the region of length $L$. Because $V_b$ is not known {\em apriori} we examine three generic situations: i) linear potential ramp along the ribbon, ii) sharp potential energy drop near the source, iii) sharp potential energy drop near drain electrode. For simplicity the potential changes in two latter cases are assumed to occur over a distance less than the lattice constant. The results presented here are for the armchair edge orientation of the ribbons. Note that similar results were obtained for zigzag edges as well.

\subsection{Uniform Ribbons}
\label{Uniform Ribbons}

Fig. \ref{fig:1}(a) shows the conductance of an ideal ribbon of uniform width for different $V_{sd}$ and bias potential profiles as a function of the Fermi energy. The ribbon length and width are chosen to be $L=100$ nm and $W=10$ nm, respectively. This corresponds to 235 unit cells and $N=80$ carbon atoms in transverse direction of the armchair edge-oriented ribbon.\cite{Dresselhaus96} Note that such ribbon sizes are readily accessible using present day fabrication techniques.\cite{Lin08, Han07} The ideal leads that carry electrons to and from the ribbon are represented by semi-infinite uniform graphene ribbons of the same width. The main effect of the source-drain bias applied to the ideal ribbon that is seen in Fig. \ref{fig:1}(a) is gradual broadening (that is symmetrical about the Dirac point $E=0$) of the transitions between the quantized conductance plateaus. This is a result of the finite energy window $eV_{sd}$ that contributes to the current in Eq. \eqref{eq:conductance}. Within that window the transmission probabilities $T$ due to states belonging to different subbands are averaged in Eq. \eqref{eq:conductance} resulting in less steep transition regions between quantization plateaus. Note that a similar effect occurs in quantum point contacts (QPCs) in the two-dimensional electron gas.\cite{qpc} The conductance for a given $V_{sd}$ is determined by availability of particular initial states for tunneling and their transmission through the channel. Only states in an energy window of width $|e|V_{sd} = \mu^s - \mu^d$ participate in transport. In the case of ballistic transmission, the integral in \eqref{eq:conductance} becomes $(n-1)(\left|e\right|V_{sd} - \epsilon^{\prime}) + n\epsilon^{\prime}$ if $\epsilon^{\prime} = \mu_s-E_n \le \left|e\right|V_{sd}$, where $E_n$ is the energy of the bottom of the highest occupied subband $n$ in the source. Thus, as $\mu_s$ changes relative to $E_n$, which might be realized by changing the Fermi energy, the conductance \eqref{eq:conductance} varies gradually between the $n-1$ and $n$ quantization plateaus making the transition region $eV_{sd}$ wide, see Fig. \ref{fig:1}(a). When $\epsilon^{\prime} > \left|e\right|V_{sd}$ the integral in \eqref{eq:conductance} will be $n$ and the conductance reaches a plateau.  Another feature of electron transport in uniform ideal graphene ribbons is the insensitivity to the details of the potential profile that is evident in Fig. \ref{fig:1}(a): Very little reduction of the conductance occurs if the smooth linear potential profile is replaced by a sharp potential drop in the ribbon; the small change might be attributed to increased electron scattering at the potential discontinuity. However, the quantized conductance values in Fig. \ref{fig:1}(a) clearly do {\em not} increase as the bias voltage increases and more subbands contribute to transport, in marked contrast to the behavior that was observed experimentally by Lin {\em et al.}\cite{Lin08} Thus the fact that the number of subbands that contribute to transport increases with increasing source drain bias is not by itself sufficient to explain the nonlinear transport phenomenon that Lin {\em et al.}\cite{Lin08} observed.

\subsection{Wide-Narrow-Wide Ribbon Geometries}
\label{Wide-Narrow-Wide}

We now consider a QPC-like structure that is more relevant to experimental devices,\cite{Han07, Lin08} where the graphene ribbon connected wider areas of graphene. Fig. \ref{fig:1}(b) shows the conductance of such a graphene ribbon QPC with the wider regions twice as wide as the constriction, $N=158$ and $N=80$ carbon atoms in cross-section, respectively. The constriction and wider regions both have armchair edges that are connected by zigzag edges at the wide-narrow boundaries, see the inset in Fig. \ref{fig:1}(b). Both of these widths support a propagating state at zero energy, which means both the wide and narrow armchair regions are metallic.\cite{Dresselhaus96} The dimensions of the constriction are similar to those of the ribbon studied in Fig. \ref{fig:1}(a). Despite this similarity the conductances in Fig. \ref{fig:1}(a) and (b) are quite different. This is due to strong electron scattering at the interfaces connecting the narrow and wide regions of the QPC ribbon. Such strong scattering was discussed in Refs. \onlinecite{Palacios07, Baranger09} and attributed to extra electron degrees of freedom on the hexagonal graphene lattice. Note that electron transport here differs substantially from that in conventional QPC's in the two-dimensional electron gas in GaAs heterostructures: No adiabatic transmission occurs in the graphene ribbon QPC's even if the interfaces between the narrow and wide regions are made atomically smooth.\cite{Palacios07} The zero bias conductance of the QPC ribbon shows strong oscillations whose envelope resembles the step-like behavior of the uniform ribbon; compare the solid and dotted lines in Fig. \ref{fig:1}(b). The oscillations result from Fabry-Perot resonant scattering at the entrance and exit of the ribbon.\cite{corrugation} They are quickly smeared by bias if $eV_{sd}$ exceeds the oscillation period and quantization steps become clearly discernible in that case. As the bias increases further the quantization steps are smoothed in a way similar to the uniform ribbon in Fig. \ref{fig:1}(a). However, two important features appear in addition. First, the conductance is suppressed more strongly, see the green dotted lines in Fig. \ref{fig:1}(b) for $\left|e\right|V_{sd}=0.025t$, $0.05t$, $0.075t$ where the linear bias ramp was assumed. In addition to the widening of the transition regions between plateaus, the plateaus themselves are lowered due to poor transmission through the QPC constriction. Second, the non-linear conductance is asymmetric about the Dirac point if the model electron potential energy in the constriction $V_b$ is such that the entire potential drop $V_{sd}$ occurs at one end of the constriction. For example, if it occurs at the end connected to the source (so that $V_b= -\frac{1}{2}\left| e\right|V_{sd}$ is ``pinned" to the drain) then the conductance curve shifts to negative Fermi energes; see the blue curve in Fig. \ref{fig:1}(b) for $\left|e\right|V_{sd}=0.025t$. The size of the shift is proportional to $V_{sd}$. The opposite shift occurs for $V_b$ pinned to the source. These asymmetries, however, become quickly damped for large $V_{sd}$ due to conductance suppression (not shown). 
\begin{figure}[t]
\includegraphics[keepaspectratio,width=\columnwidth]{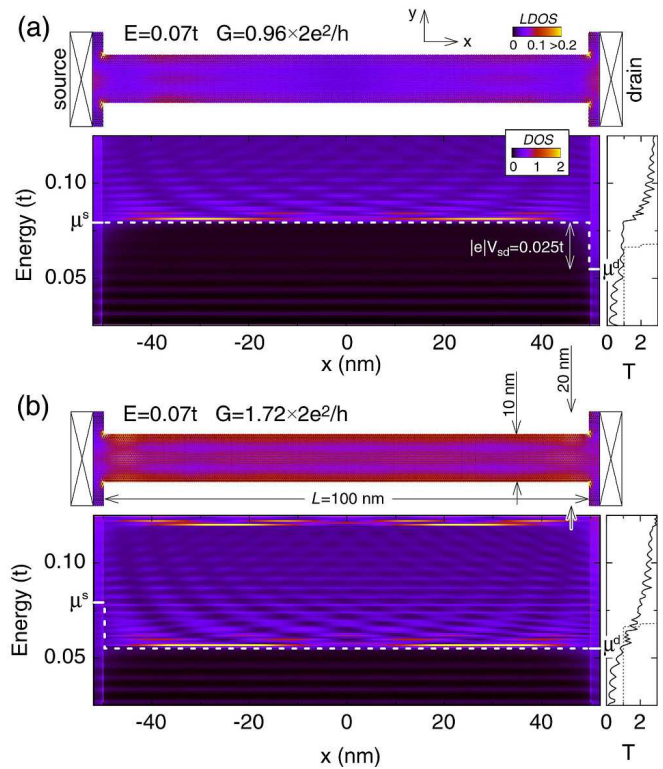}
\caption{(color online) Representative LDOS and DOS for the QPC ribbons for the Fermi energy $E_\mathrm{F}=0.07t$ and source-drain bias $V_{sd}=0.025t/|e|$. DOS was calculated by integration across the ribbon. The constriction potential equals the source (a) or drain (b) potential. The small outsets on the right show the transmission coefficient. }
\label{fig:2}
\end{figure}

The conductance asymmetry in Fig. \ref{fig:1}(b) that is triggered by electron scattering at the ends of the QPC constriction can be understood by considering the DOS plots in Fig. \ref{fig:2}. The figure presents cases of source (a) and drain (b) ``pinned" constriction potentials for a representative Fermi energy $E_\mathrm{F}=0.07t$. Conduction through the QPC structure is constrained by the constriction, not by the electrodes where there are more transport states at every energy. If the constriction potential equals that in the source, Fig. \ref{fig:2}(a), electrons from the second and third subband in the leads have no corresponding available states in the constriction and transmission in the energy range $eV_{sd}$ surrounding $E_\mathrm{F}$ is damped. The total transmission is close to 1 due to the first subband state that still propagates freely in the constriction at $E=0.07t$. Inspection of the LDOS in the  upper panel of Fig. \ref{fig:2}(a) reveals low amplitudes inside the constriction. If $V_b= -\left| e\right| V_{sd}/2$ as in Fig. \ref{fig:2}(b) the second and third subbands in the constriction become available for electron transmission to the drain electrode. The transmission is, however, not perfect due to state mismatch at the interfaces.\cite{Palacios07, Baranger09} The DOS inside the constriction shows multiple resonant peaks due to the Fabry-Perot interference. Each peak is due to a resonant state available for propagation with enhanced transmission $T$ at the corresponding energy. The LDOS at the Fermi energy for the drain-pinned bias potential is much higher than for the source-pinned potential; compare the upper plots in Fig. \ref{fig:2}(a) and (b). The larger LDOS is a manifestation of the higher probability to find an electron in the constriction. Note that by far the strongest electron localization occurs at the corners of the structure regardless of the potential distribution. If the Fermi energy changes sign and the charge carriers become holes the availability of transport states inside the constriction reverses. This results in the conductance curve being shifted in energy by an amount proportional to $V_{sd}$ relative to its position at zero bias. The conductance shows no asymmetry about the Dirac point for the linear ramp potential because of approximately similar availability of transport states inside the constriction for the Fermi energies of opposite sign. 

In Fig. \ref{fig:1}(b), for the QPC models with the potential drop at one end of the constriction, the conductance increases with increasing source drain bias in a limited range of low bias values for some values of the Fermi energy. However, these conductance increases occur in the transition regions between the quantized conductance plateaus. The quantized conductance values on the plateaus themselves do not increase with increasing bias although the Fermi energy (or gate voltage) ranges in which the plateaus occur shift as the bias increases. This behavior is again qualitatively different than that observed by Lin {\em et al.}\cite{Lin08} who reported the {\em quantized} conductance values increasing with increasing bias without the ranges of gate voltage in which the quantized plateaus occurred changing significantly. We conclude that the observed nonlinear transport phenomenon\cite{Lin08} is unlikely to be a property of ballistic graphene ribbon QPC's. We will show next that electron scattering due to disorder can give rise to this effect.

\section{Uniform ribbons with defects}
\label{defects}
\subsection{Results}
\label{Results}

\begin{figure}[t]
\includegraphics[keepaspectratio,width=\columnwidth]{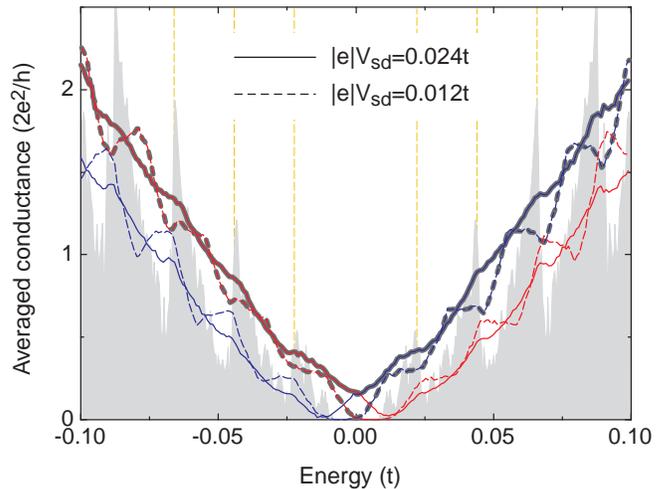}
\caption{(color online) The averaged conductance vs. Fermi energy for 30 nm wide and 500 nm long graphene ribbon with $p=10^{-4}$ bulk vacancies. The dashed and solid lines correspond to $\left| e \right| V_{sd}=0.012t$ and $0.024t$, respectively. Note that valley degenerate subbands are separated by $\sim0.024t$ for $30$ nm wide ribbon; the subband positions are marked by vertical dashed yellow lines. The blue and red lines show results for source and drain pinned constriction potentials. The gray filled area denote zero bias conductance. The branches of the conductance curves that resemble the experimental non-linear conductance data of Lin {\em et al.}\cite{Lin08} are outlined by the thick dark gray dashed and solid lines. Averaging is performed over 10 samples with different realizations of the bulk vacancies.}
\label{fig:3}
\end{figure}

We calculated the averaged conductances for uniform ribbons with bulk vacancy disorder. Representative results are shown in Fig. \ref{fig:3}. The ribbon width was chosen to be $W=30$ nm, the same as that of the experimental samples in Ref. \onlinecite{Lin08}. The presence of the bulk vacancies leads to enhanced electron backscattering near subband edges that in turn suppresses the conduction equally for all subbands resulting at low bias in conductance steps of equal height\cite{disorder} similar to those observed by Lin {\em et al.}\cite{Lin08} Edge disorder as well as long-range potentials are probably present in the experimental samples\cite{Lin08} but have been shown\cite{disorder} not to be of prime importance for the observed conductance quantization and we, therefore, concentrate here on scattering by bulk vacancies. It is worth noting that adsorbed H atoms cause electron scattering and conductance quantization that are very similar to those due to bulk vacancies.\cite{adsorbate}

Figure \ref{fig:3} shows the calculated non-linear conductances for the ribbon with bulk vacancies when the bias potential in the ribbon is pinned to the source or drain potential, i.e. $V_b=+ \left| e\right| V_{sd}/2$ or $V_b=- \left| e\right| V_{sd}/2$. The plots are for two values of  $\left| e \right| V_{sd}=0.012t$ and $0.024t$ that differ by half of a subband energy separation. When $\left| e \right| V_{sd}$ increases from $0.012t$ and $0.024t$ the conductance plateaus shift upward (with some smoothing) by half of the conductance quantization step for electron transport if $V_b=- \left| e\right| V_{sd}/2$ and for hole transport if $V_b=+ \left| e\right| V_{sd}/2$. The corresponding branches of the conductance curves are outlined by the thick dark gray dashed and solid lines in Fig. \ref{fig:3} and resemble the experimental results in Fig.4 of Ref. \onlinecite{Lin08}. That is, in the present model, the conductance plateaus move to higher conductance values with increasing bias as in the experiment of Lin {\em et al.}\cite{Lin08} if the potential in the ribbon is pinned to that in the drain for electron transport and to that in the source for hole transport. By contrast, if a linear ramp potential in the ribbon is assumed, the quantized conductance plateaus are found not to shift to higher conductance values with increasing source-drain bias. Instead the transitions between them broaden until the plateaus disappear.

\subsection{Discussion}
\label{Discussion}

The model ribbon potential profiles that we have considered above have been phenomenologically motivated. However, we shall now argue that it is physically reasonable that $V_b$ is approximately uniform throughout most of the ribbon and that it changes from being pinned to the drain to being pinned to the source when the Fermi energy crosses from above to below the Dirac point: Calculations of the potential profiles in idealized models of short (a few nanometers long) gated graphene ribbons have indicated that due to screening most of the potential drop in the ribbon that is due to the applied source-drain bias occurs near the contacts.\cite{Cheraghchi10}  The characteristic length scale for such potential variations has been estimated\cite{QZhang08} to be $\lambda = \sqrt{(\epsilon_{GNR}/\epsilon_{OX})t_{GNR}t_{OX}}$ where $\epsilon_{GNR}$ and $\epsilon_{OX}$ are the graphene ribbon and gate oxide dielectric constants and $t_{GNR}$ and $t_{OX}$ are the graphene ribbon and gate oxide thicknesses. For the ribbon of Lin {\em et al.}\cite{Lin08} this yields $\lambda \sim 10$ nm which is much smaller than the ribbon length of 1.7 $\mu$m in their non-linear conductance experiment. This suggests that our assumption that $V_b$ is constant throughout most of the ribbon is justified. We also note that since the measured conductances of the ribbons of Lin {\em et al.}\cite{Lin08} were much smaller that $2e^2/h$ most of the charge carriers entering the ribbon were reflected back into the electrode from which they came without penetrating deeply into the ribbon. Thus most of the charging of the ribbon by the carriers and hence most of the variation of the electrostatic potential in the ribbon should have occured near the end of the ribbon where the carriers enter, i.e., near the source electrode for electrons and the drain for holes. Consequently throughout most of the ribbon the potential is expected to be pinned at the drain potential for Fermi levels above the Dirac point (electron carriers) and at the source potential for Fermi levels below the Dirac point (hole carriers), in agreement with the model potential profiles that we have shown above to yield behavior of the nonlinear quantized conductances resembling that observed experimentally by  Lin {\em et al.}\cite{Lin08} 
  
The nonlinear conductance in the experiment\cite{Lin08} exhibited plateaus at $V_{sd}=100$ mV with conductances higher than those for $V_{sd}=50$ mV by roughly one unit of the quantized conductance but with similar plateau widths. Note that 50 mV is energy separation between valley-degenerate subbands. However, our results show the quantization steps becoming smoother as the bias increases. This discrepancy might be attributed to the effect of electron-electron interactions. At the same time the experimental paper\cite{Lin08} presented just a single measurement for a single sample. Figure \ref{fig:3} shows, by contrast, a statistically averaged result. Therefore, sample-specific conductance quantization should not be ruled out when experimental data in Ref. \onlinecite{Lin08} is analyzed.

\section{Conclusions}
\label{Conclusions}

In conclusion, we have studied conductance  quantization in graphene ribbons in the non-linear transport regime. We found that with increasing source-drain bias the transitions between the quantized conductance plateaus broaden. We also found that the conductance plateaus show significant asymmetry between the electron and hole branches if the potential in the channel is pinned either to the source or to the drain electrode potential and strong electron (hole) scattering occurs. This scattering may occur at the ends of a uniform ballistic ribbon that connects wider regions of graphene or may be due to defects in the interior of a ribbon. We argue that in a ribbon with strong defect scattering the potential in the ribbon should be pinned to the drain potential for electron transport and to the source potential for hole transport. In that case the electron-hole symmetry of the conductance is recovered and our results explain the upward shift of the conductance plateaus with increasing bias between source and drain observed experimentally by Lin {\em et al.}\cite{Lin08} for both electron and hole transport. Further theoretical study that accounts for electron-electron interactions is required and more experimental data are needed to better understand non-linear electron transport in the graphene ribbons.   

\begin{acknowledgments}
This work was supported by NSERC, CIFAR and WestGrid. 
\end{acknowledgments}


\begin{thebibliography}{99}

\bibitem{Lin08} Yu-Ming Lin, V. Perebeinos, Zhihong Chen, and Ph. Avouris, Phys. Rev. B \textbf{78}, 161409(R) (2008).

\bibitem{Kirczenow}  For a review see G. Kirczenow, in The Oxford Handbook of
Nanoscience and Technology,Volume I: Basic Aspects, Chapter 4.6, edited by A. V.
Narlikar and Y. Y. Fu, Oxford University Press, U.K. (2010).

\bibitem{disorder} S. Ihnatsenka and G. Kirczenow, Phys. Rev. B \textbf{80}, 201407 (2009).

\bibitem{adsorbate} G. Kirczenow and S. Ihnatsenka, arXiv:1008.4110; S. Ihnatsenka and G. Kirczenow, arXiv:1008.4111 (unpublished).

\bibitem{Lian2010} C. Lian, K. Tahy, T. Fang, G. Li, H. G. Xing, and D. Jena, Appl. Phys. Lett. {\bf 96}, 103109 (2010).

\bibitem{Tombros2011} N. Tombros, A. Veligura, J. Junesch, M. H. D. Guimar\"{a}es, I. J. V. Marun, H. T. Jonkman and B. J. van Wees, arXiv:1102.0434v1.

\bibitem{Reich02} S. Reich, J. Maultzsch, C. Thomsen, and P. Ordej\'{o}n, Phys. Rev. B \textbf{66}, 035412 (2002).

\bibitem{Evaldsson08} M. Evaldsson, I. V. Zozoulenko, Hengyi Xu and T. Heinzel, Phys. Rev. B \textbf{78}, 161407(R) (2008).

\bibitem{Mucciolo09} E. R. Mucciolo, A. H. Castro Neto, and C. H. Lewenkopf, Phys. Rev. B \textbf{79}, 075407 (2009).

\bibitem{Areshkin07} D. Areshkin, D. Gunlycke, and C. T. White, Nano Lett. \textbf{7}, 204 (2007).

\bibitem{Davies_book} J. Davies, \textit{The Physics of Low-Dimensional Semiconductors}, (Cambridge University Press, Cambridge, 1998).

\bibitem{Igor08} Hengyi Xu, T. Heinzel, M. Evaldsson, and I. V. Zozoulenko, Phys. Rev. B \textbf{77}, 245401 (2008).

\bibitem{Datta_book} S. Datta, \textit{Electronic Transport in Mesoscopic Systems}, (Cambridge University Press, Cambridge, 1997).

\bibitem{Dresselhaus96} K. Nakada, M. Fujita, G. Dresselhaus and M. S. Dresselhaus, Phys. Rev. B \textbf{54}, 17954 (1996).

\bibitem{Han07} M. Y. Han, B. \"{O}zyilmaz, Y. Zhang, and P. Kim, Phys. Rev. Lett. \textbf{98}, 206805 (2007); F. Molitor, A. Jacobsen, C. Stampfer, J. G\"{u}ttinger, T. Ihn, and K. Ensslin, Phys. Rev. B 79, 075426 (2009); C. Stampfer, J. G\"{u}ttinger, S. Hellm\"{u}ller, F. Molitor, K. Ensslin, and T. Ihn, Phys. Rev. Lett. \textbf{102}, 056403 (2009).

\bibitem{qpc} S. Ihnatsenka and I. V. Zozoulenko, Phys. Rev. B \textbf{79},  235313 (2009).

\bibitem{Palacios07} F. Mu\~{n}oz-Rojas, D. Jacob, J. Fern\`{a}ndez-Rossier, and J. J. Palacios, Phys. Rev. B \textbf{74}, 195417 (2006).

\bibitem{corrugation} S. Ihnatsenka, I. V. Zozoulenko and G. Kirczenow, Phys. Rev. B \textbf{80}, 155415 (2009); G. Kirczenow, Solid State Commun. \textbf{68}, 715 (1988), Phys.  Rev. B. \textbf{39}, 10452 (1989).

\bibitem{Baranger09} J. Wurm, M. Wimmer, I. Adagideli, K. Richter and H. U. Baranger, New J. Phys. \textbf{11}, 095022 (2009). 

\bibitem{Cheraghchi10} H. Cheraghchi and H. Esmailzade, Nanotechnology {\bf 21}, 205306 (2010) 

\bibitem{QZhang08}Q. Zhang, T. Fang, H. Xing, A. Seabaugh and D. Jena, IEEE Electron Device Lett., {\bf 29}, 1344 (2008).

\end{thebibliography}
\end{document}